\documentclass[prb,preprint]{revtex4} 
\usepackage{epstopdf}
\usepackage{rotating}
\usepackage{amsmath}

\begin{document} 

\title{How superluminal motion can lead to backward time travel}

\author{Robert J. Nemiroff}
\email{nemiroff@mtu.edu}
\author{David M. Russell}
\email{dmrussel@mtu.edu}
\affiliation{Michigan Technological University, Department of Physics, 
1400 Townsend Drive, Houghton, Michigan 49931}


\begin{abstract}
It is commonly asserted that superluminal particle motion can enable backward time travel, but little has been written providing details. It is shown here that the simplest example of a ``closed loop" event -- a twin paradox scenario where a single spaceship both traveling out and returning back superluminally -- does {\it not} result in that ship straightforwardly returning to its starting point before it left. However, a more complicated scenario -- one where the superluminal ship first arrives at an intermediate destination moving subluminally -- can result in backwards time travel. This intermediate step might seem physically inconsequential but is shown to break Lorentz-invariance and be oddly tied to the sudden creation of a pair of spacecraft, one of which remains and one of which annihilates with the original spacecraft.
\end{abstract}

\maketitle

\section{INTRODUCTION}

Objects traveling faster than light are discouraged by popular convention in Einstein's Special Theory of Relativity \citep{Einstein1905}, which provides a profound, comprehensive, and experimentally verified description of particle trajectories and kinematics at subluminal speeds. Nevertheless, the vast distance to neighboring stars has caused superluminal speeds to continue to be discussed in popular venues. \citep{StarTrek1964} To be clear, this work is {\it not} advocating that faster than light speeds for material particles are possible. Rather, the present work takes superluminal particle speeds as a premise to show how closed-loop backward time travel arises in a specific simple scenario.

Physics literature has indicated for many years that superluminal speeds can correspond to backward time travel. \citep{Tolman1917} Such claims are pervasive enough to have become common knowledge, as exemplified by a famous limerick published in 1923: ``There was a young lady named Bright, Whose speed was far faster than light; She set out one day, In a relative way, And returned on the previous night".\citep{Buller1923} 

The possibly of closed-loop time travel within the context of special relativity was later mentioned explicitly in 1927 by Reichenbach. \citep{Reichenbach1927} A prominent discussion on the physics of particles moving superluminally within the realm of special relativity was given in 1962 by Bilaniuk Deshpande, and Sudarshan. \citep{Bilaniuk1962} The term ``tachyon" was first coined for faster than light particles by Feinberg \citep{Feinberg1967} who also derived transformation equations for superluminal particles. Tachyonic speeds have been suggested multiple times in the physics literature to address different concerns, for example being convolved with quantum mechanics to create pervasive fields \citep{Feinberg1967}, and to explain consistent results between two separated detectors in quantum entanglement experiments. \citep{Einstein1935}, \citep{Bell1966} 

The reality of particle tachyons or any local faster-than-light communication mechanism is controversial, at best. Accelerating any material particle from below light speed to the speed of light leads to a divergence in the particle's energy, a physical impossibility. For $v > c$, The Lorentz-FitzGerald Contraction \citep{Lorentz1892}, \citep{FitzGerald1889} $\sqrt{1 - v^2/c^2}$ becomes imaginary, leading to relative quantities like mass, distance, and time becoming ill-defined, classically. Simple tachyonic wavefunctions in quantum mechanics either admit only subluminal or non-localizable solutions. \citep{Chase1993} Experimental reports of particles moving faster than light have all been followed by skeptical inquiries or subsequent retractions. \citep{Opera2011}, \citep{Opera2012} Were tachyonic communications to enable communications backward in time, violations in causality seem to result, a prominent example of which is the Tachyonic Anti-telephone. \citep{Benford1970} For this reason superluminal communication and backward time travel are thought to be impossible. Experimentally, a recent search of Internet databases for ``unknowable-at-the-time" information that might indicate the possibility of backward time travel came up empty. \citep{NemWil2014}

Conversely, the existence of superluminal speeds for phase velocities and illumination fronts that do not carry mass or information are well established. \citep{Griffiths1994} The ability of superluminal illumination fronts to show pair creation and annihilation events was mentioned by Cavaliere et al. \citep{Cavaliere1971} and analyzed in detail by Nemiroff \citep{Nemiroff2015} and Zhong \& Nemiroff \citep{Zhong2016}. 

The possibility that a material object could undergo a real pair creation and subsequent annihilation event was mentioned in 1962 by Putnam \citep{Putnam1962} including the possibility of pair events with regard to backward time travel. However, Putnam's treatment was conceptual, gave no mathematical details, and the concept of a closed loop was not considered. In 2005 Mermin \citep{Mermin2005} noted such behavior for an object moving subluminally but with an intermittent period of superluminal motion, reporting that such pair events would only be evident in some inertial frames. Mermin also never considered a closed loop event.

There appears to be no detailed treatment, however, showing {\it how} superluminal speeds lead to ``closed-loop" backward time travel: a material observer returning to a previously occupied location at an earlier time. Treatments generally stop after showing that faster than light objects can be seen to create negative time intervals for relatively subluminal inertial observers. \citep{Tolman1917} To fill this void, the standard velocity addition formula of special relativity is here applied in the superluminal domain to show that closed-loop backward time travel can, in specific circumstances, be recovered -- but perhaps in a surprising way. This scenario can also be considered a didactic and conceptual extension of the famous ``twin paradox" \citep{Einstein1905} to superluminal speeds.

\section{OUT AND BACK AGAIN}

The scenario explored here is extremely simple: an object goes out and comes back again. The return trip is important to ensure a ``closed-loop". Therefore, the scenario described can be thought of as an extension of the famous twin paradox to superluminal speeds. For the sake of clarity, to promote interest, and to place distances on scales where temporal effects correspond with common human time scales, the initial launching location will be called ``Earth", which can also be thought of as representing the twin that stays at home. The ballistic projectile will be referred to as a ``spaceship", which can also be thought of as representing the twin the travels away and then comes back. The turnaround location will be referred to as a ``planet". Furthermore, an example where the distance scale is on the order of light-years will be described concurrently.

The following conventions are observed. In general, unless stated otherwise, all times and distances will be given in the inertial frame of the Earth and from the location of Earth. All relative motion for the spaceship will take place in the line connecting the Earth to the planet, here defined as the $x$ axis. All velocities are assumed constant. The planet is assumed moving away from the Earth at the subluminal speed $u$ as measured in the Earth's inertial frame. Times are given by the variable $t$, and the standard time when the spaceship is scripted to leave Earth is set to $t = 0$. At this time the spaceship leaves Earth from a location called the Launch Pad, and aims to return to an Earth location called the Landing Pad. The distance on Earth between the Launch Pad and the Landing Pad is considered negligible. Velocities away from the Earth are considered positively valued, and velocities toward the Earth are considered negatively valued. The outbound velocity of the ship relative to the Earth is given by $v$, the return velocity of the ship relative to the planet is given by $-v$, and the return velocity of the ship relative to the Earth is designated $w$. Keeping both spaceship speeds at magnitude $v$ is a useful didactic simplification that demonstrates the logic of a much larger set of event sequences when the outgoing and incoming spaceship speeds are decoupled. 

At time $t = 0$, the planet is designated to be a distance $x_{po}$ and moving at positive velocity $u$ away from the Earth, with respect to the Earth. Therefore, at time $t$ the distance between Earth and the planet is simply 
 \begin{equation}
  x_p (t) = x_{po} + u t .
 \end{equation}

Similarly, at time $t=0$, the spacecraft leaves Earth from the Launch Pad. After launch and before reaching the planet, the distance between Earth and the spaceship is
 \begin{equation} \label{xout}
  x_{out} (t) = v t .
 \end{equation}
The x-coordinate will usually describe the spaceship and so, when it does, no subscript will be appended.

The spacecraft reaches the planet at the time when $x(t) = x_p(t)$. Combining the above two equations shows that the amount of time it takes for the spacecraft to reach the planet is
 \begin{equation} \label{tout}
 t_{out} = \Delta t_{out} = { x_{po} \over v - u } .
 \end{equation}
The distance to both the spacecraft and the planet at this time is
 \begin{equation} \label{xturn}
 x_{turn} = { v x_{po} \over v - u} .
 \end{equation}

The spaceship turns around at the planet. For this calculation, the turnaround is considered instantaneous but the important point is that the turnaround duration is small compared to other time scales involved. After turnaround, the velocity of the spaceship relative to the Earth is
 \begin{equation} \label{weq}
 w = { u + v  \over 1 + u v / c^2 } ,
\end{equation}
where $u$ is measured relative to the Earth but $v$ is measured relative to the planet. Only in Eq. (\ref{weq}) is $v$ negative as it describes the spaceship returning back to Earth -- in all other equations $v$ is to be considered positive as it refers to the speed of the ship leaving Earth. Note that Eq. (\ref{weq}) is the standard equation of velocity addition in special relativity and one that has been consistently invoked even when superliminal speeds are assumed. \citep{Mermin2005}, \citep{Hill2012} Because of the centrality of this equation to physics, it is retained here in its classic form.

After the spaceship leaves the planet, its distance from the Earth is given by
 \begin{equation} \label{xback}
  x_{back} (t) = x_{turn} + w (t - t_{out}) 
               = { v x_{po} \over v - u }  + { (u - v) (t - x_{po} / (v - u)) \over 1 - uv/c^2 } .
 \end{equation}
The scenario is defined so that once the spaceship returns to Earth, it lands on the Landing Pad and stays there. The spacecraft can only move between the Earth and the planet -- it does not go past the Earth to negative $x$ values.

The time it takes for the spaceship to return back to the Earth from the planet is  
 \begin{equation} \label{tback0}
 \Delta t_{back} = - x_{turn} / w .
 \end{equation}
The negative sign leading this equation is necessary to make the amount of return time positive when $w$ is negative. Writing $\Delta t_{back}$ in terms of (positively valued) $u$, $v$, and $x_{po}$ yields
 \begin{equation} \label{tback}
 \Delta t_{back} = { v x_{po} (1 - u v / c^2) \over (u - v)^2 } .
 \end{equation}

The total time that the ship takes for this trip is 
 \begin{equation} \label{ttot}
 \Delta t_{tot} = t_{tot} = t_{out} + \Delta t_{back}
 = { x_{po} ( 2 v - u - u v^2 / c^2)  \over (v - u)^2 } ,
 \end{equation}
where, again, all interior speeds are defined as being positively valued and $v > u$. Were the planet stationary with respect to Earth, parameterized by $u = 0$ then $\Delta t_{out} = \Delta t_{back} = \Delta t_{tot} / 2 = x_{po} / v$ which agrees with the non-relativistic classical limit, no matter the (positive) value of $v$. 

It is also of interest to track how long it takes light signals to go from the spaceship to Earth, as measured on Earth. The time after launch that an Earth observer sees the spaceship at position $x(t)$ will be labeled $\tau(t)$. Since light moves at $c$ in any frame, then Earth observers will see the outbound spaceship at position $x(t)$ at time
 \begin{equation} \label{iout}
 \tau_{out} (t) = x_{out} (t) / v + x_{out} (t) / c ,
 \end{equation}
where the first term is the time it takes for the spacecraft to reach the given position, and the second term is the time it takes for light to go from this position back to Earth. The time that Earth observers will see the spacecraft reach the planet is
 \begin{equation} \label{iturn}
 \tau_{turn} = x_{turn} / v + x_{turn} / c.  
 \end{equation}

During the spaceship's return back to Earth, Earth observers will see the spaceship at $x_{back} (t)$ such that  
 \begin{equation} \label{iback} 
 \tau_{back} (t) = x_{turn} / v 
                 + (x_{back}(t) - x_{turn}) / w 
                 + x_{back} (t) / c.
 \end{equation}
The first term is the time it takes for the spacecraft to reach the planet, the second term is the time it takes for the spacecraft to go from the planet to intermediate position $x_{back}(t)$, and the third term is the time it takes for light to reach Earth from position $x_{back} (t)$. 

A series of threshold $v$ values occur, which will be reviewed here in terms of increasing magnitude. 

\subsection{Threshold Speed: $v = u$}

The first threshold speed explored is $v = u$. Below this speed, the spaceship is moving too slowly to reach the planet. When $v = u$, the spaceship has the same outward speed as the planet and only reaches the planet after an infinite time has passed. This is shown by the denominators going to zero in Eqs. (\ref{tout}, \ref{tback}, and \ref{ttot}). Earth observers will see both the spaceship and the planet moving away in tandem forever.

\subsection{Spaceship speeds $u < v < c$}

When $u < v < c$, and when both the spaceship and the planet are moving much less than $c$, then Earth observers see the spaceship move out to the planet and return back to Earth in a normal fashion that is expected classically.

When $u < v < c$ generally, then $\Delta t_{back} > \Delta t_{out}$. This results from the magnitude of the spacecraft's speed coming back to Earth, $w$, being less than the magnitude of the spacecraft's speed going out to the planet, $v$, even though the distance traveled by the spacecraft is the same in both cases: $x_{turn}$. In this speed range, the occurrence of events in the Earth's inertial frame proceeds as expected in non-relativistic classical physics. As tracked from the Earth, the spaceship simply goes out to the planet, turns around, and returns.

\begin{figure}[h]
\includegraphics[angle=90, width=18cm]{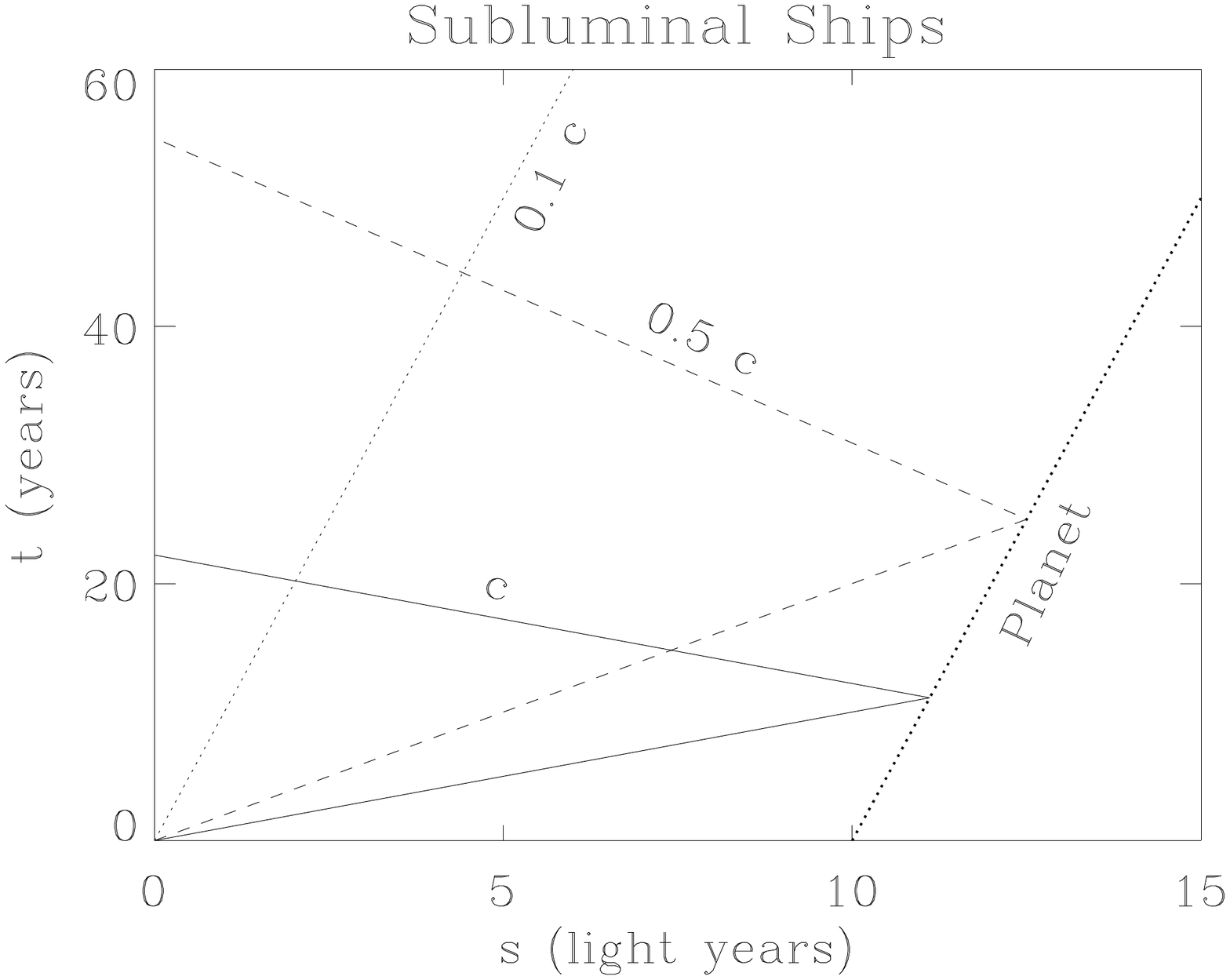}
\caption{A spacetime diagram is shown for the journeys taken by the subluminal ships described in the text. Time is plotted against distance traveled, both measured in the frame of the Earth. Earth remains at $(0,0)$. The world line of the planet continually receding from the Earth at $v_p = 0.1 c$, in the Earth frame, is depicted by the dark dotted line on the right. A space ship traveling out at $v = 0.1 c$ in the Earth frame is shown by the light dotted line. This ship never reaches the planet.  A ship traveling out at $v = 0.5 c$ in the Earth frame, reaching the planet, and returning to Earth at $v = 0.5 c$ in the planet's frame, is shown by the dashed line. A ship similarly traveling at $v = c$ is shown by the solid line.}
\label{superslow}
\end{figure}

For clarity, a series of specific numerical examples are given, with values echoed in Table \ref{table1}, and world lines depicted in the Minkowski spacetime diagrams of Figures \ref{superslow} and \ref{superfast}. In all of these examples, the planet starts at distance $x_{po} = 10$ light years from Earth when $t = 0$, and the planet's speed away from the Earth is $u = 0.1 c$. For the speed range being investigated in this sub-section, the spaceship has speed $v = 0.5 c$. Then by Eq. (\ref{xturn}) the spaceship reaches the planet when both are $x_{turn} = 12.50$ light years from Earth. The duration of this outbound leg is given by Eq. (\ref{tout}) as $\Delta t_{out} = 25.00$ years. For clarity, the speed of the ship's return is computed from Eq. (\ref{weq}) as $w \sim -0.4211 c$, to four significant digits. The duration of the ship's return back to Earth is given by Eqs. (\ref{tback0}, \ref{tback}) as $\Delta t_{back} \sim 29.69$ years. The total time that the spaceship is away is the addition of the ``out" time and the ``back" time, which from Eq. (\ref{ttot}) is $\Delta t_{tot} = 54.69$ years. The world line of this ship's trip is given by the dashed line in Figure 1.

Due to the finite speed of light, Earth observers perceive the spacecraft as arriving at the planet only after the equations indicate that it has already started back toward Earth. The closer $v$ is to $c$, the closer the spacecraft is to the Earth, as defined by Eq. (\ref{iturn}), when Earth observers see the spaceship arrive at the planet.

In the concurrent example of $v = 0.5 c$, the spaceship reaches the planet at time $t = 25.00$ years, but light from this event does not reach Earth until the time given by Eq. (\ref{iturn}) -- $\tau_{turn} = 37.50$ years, well after the spaceship has actually left the planet. However, the ship is first seen on Earth to arrive back at Earth when it actually arrives back on Earth -- 54.69 years after it left.

\subsection{Threshold Speed: $v = c$}

The next threshold value for the spaceships outbound speed explored is $v = c$.  Generally when $v = c$, however, then not only is the spaceships speed $c$ relative to Earth on the way out, but it is $c$ relative to the {\it planet} on the way out, it is $-c$ relative to Earth on the way back, and it is $-c$ relative to the planet on the way back. Here $\Delta t_{out} = \Delta t_{back}$: the spaceship takes the same amount of time to reach the planet as it does to return.

In the example, the only quantity changed is the speed $v$ of the spaceship. At $v = c$, the spaceship catches up to the planet at $x_{turn} = 11.11$ light years at time $t_{out} = 11.11$ years after launch. The speed of return is $w = -c$. The time it takes for the ship to return is $\Delta t_{back} = 11.11$ years, meaning the total time for the trip, as measured on Earth, is $\Delta t_{tot} = 22.22$ years.

What Earth observers see, in general, is quite different from the classical non-relativistic cases. Neglecting redshifting effects, the spaceship would appear to travel out to the planet normally, but would arrive back on Earth at the same time that the spacecraft appears to arrive at the planet. In fact, light from the entire journey back to Earth would arrive at the same time the spacecraft itself arrived back on Earth. This is because as perceived from Earth, the spacecraft, returning at speed $c$, rides alongside all of the photons it releases toward Earth on the way back.

In the specific example, the $v = c$ spaceship is seen to arrive back on Earth when $x_{back}(t_{tot}) = 0$, which Eq. (\ref{iback}) gives as $\tau_{back} = \tau_{turn} =$ 22.22 years. The solid lines in Figures \ref{superslow} and \ref{superfast} depict the world lines of this ship.

\subsection{Spaceship speeds $c < v < c^2/u$}

To explore the question as to how faster-than-light motion can lead to backward time travel, it is now supposed that superluminal speeds are possible for material spaceships. In general, as greater spacecraft speeds $v$ are considered in the range $c < v < c^2/u$, the magnitude of the speed of the spacecraft's return $w$ increases without bound. Here, in general, $\Delta t_{back} < \Delta t_{out}$. 

In the specific example, the spaceship speed is now taken to be $v = 5 c$. The spaceship catches up with the planet at $x_{turn} = 10.20$ light years at time $t_{out} = 2.041$ years after launch. The return speed is $w = -9.800 c$ so that it takes the spaceship $\Delta t_{back} = 1.041$ years to get back to Earth. Earth receives the spaceship back after 3.082 years away. 

What Earth observers see, in general, is perhaps surprising, as tracked by Eqs. (\ref{iout}) and (\ref{iback}). First the spacecraft appears to leave for the planet as normal. Next, however, two additional images of the spacecraft appear on Earth on the Landing Pad, one of which stays on the Landing Pad, while the other image immediately appears to leave for the planet. The underlying reason for these strange apparitions is that spacecraft itself returns to Earth {\it before} two images of the spacecraft return to Earth. Therefore, after the spacecraft returns, the Earth observer sees not only the returned spacecraft, but an image of the spacecraft on the way out, and an image of the spacecraft on the way back, all simultaneously. 

In the specific example, an image of the $v = 5 c$ spacecraft is seen moving toward the planet and arriving, as determined by Eq. (\ref{iturn}) at $\tau_{turn} = 12.24$ years, even though the spacecraft itself arrived back on Earth earlier -- after only 3.082 years. The light dotted line in Figure \ref{superfast} describes the world line for this trip's journey.

It is particularly illuminating to consider what is visible from Earth five years after the spacecraft left the Launch Pad, after the actual spacecraft has arrived back on Earth but before the spacecraft appears to have arrived at the planet. First, assuming the returned spacecraft has remained on the Landing Pad, there is the image of the returned spacecraft remaining on the Landing Pad. Next, an image of the spacecraft going out to the planet is visible back on Earth. Focusing on this image of the outbound craft, one can solve Eq.(\ref{iout}) for $x_{out}$ to find that $x_{out} = 4.167$ light years from Earth when this image of the outbound ship arrives back on Earth. Last, a third image -- an image of the spacecraft on its return back from the planet -- is simultaneously visible back on Earth. At the five-year mark, one can solve Eq. (\ref{iback}) to find that $x_{back}(5 \ {\rm years}) = 2.135$ light years. Therefore, Earth sees this third image as the spaceship is returning to Earth, but still 2.135 light years distant. 

To recap, five years after leaving the Launch Pad, three images of the spacecraft are visible on Earth. One image is emitted by the spacecraft as it remains sitting on the Landing Pad after its return, another image is emitted by the spacecraft on its way to the planet, and a third image is emitted by the spacecraft on its way back from the planet.

It is further illuminating to check on the spacecraft images still visible on Earth {\it eight} years after it left the Launch Pad, three years after the previous check. At eight years, assuming the spaceship remains on the Landing Pad, an image of the returned spaceship remains visible on the Landing Pad. Solving Eq. (\ref{iout}) for $x_{out}$ now shows the ``outbound image" of the ship at location 6.667 light years from Earth, further out than it was before. The image of the returning spaceship shows from Eq. (\ref{iback}) the ship at a distance 5.477 light years from Earth. This might appear odd as the return image arriving at Earth eight years out shows the ship as further away from Earth -- not closer -- than the image of the returning craft that arrived after five years. It therefore appears that this return spaceship is moving backward in time, as seen from Earth.

For clarity, to recap again, eight years after leaving the Launch Pad, three images of the spacecraft remain visible from Earth: one image on the Landing Pad, one image on the way out, and one image on the way back. Both the outbound image and the return image show the ship appearing further away after year eight than at year five. 

Right at the time the spaceship returns to Earth, the number of spaceship images visible on Earth jumps from one to three. Before this, Eq. (\ref{iback}) shows that both the image of the spaceship on the Landing Pad and an image of the spaceship returning to the Landing Pad have yet to reach Earth. Therefore for $v$ in this interval, the spaceship reaching the Landing Pad marks an image pair creation event. 

Similarly, when the spacecraft reaches the planet, both the outbound and the return images of this event arrive back at Earth simultaneously, as can be seen from Eqs. (\ref{iout}) and (\ref{iback}). Because no further images of the spacecraft going out or returning exist, these images then both disappear, leaving only the spacecraft image on the Landing Pad. This disappearance is an image pair annihilation event. These image pair events are conceptually similar to spot pair events seen for non-material illumination fronts moving superluminally. \citep{Nemiroff2015} Image events are entirely perceptual -- the actual location of the spaceship is given at any time by Eqs. (\ref{xout}) and (\ref{xback}). Observers at other vantage points -- or in other inertial frames -- may see things differently, including possible different relative timings of image pair creation and annihilation events. Also, in this velocity range, since the location of the spacecraft $x(t)$ in the Earth frame is unique, it is clear that only one spaceship ever exists at any given time.

\subsection{Threshold Speed: $v = c^2/u$}

At the threshold speed $v = c^2/u$, the speed of return $w$ of the spacecraft to Earth diverges as the denominator of Eq. (\ref{weq}) goes to zero. Although slight variations of $v$ below and above this threshold speed will yield $w$ divergences to positive or negative infinity, the negative infinity realization will be discussed. At formally infinite return speed $w$, the time it takes for the spacecraft to return to Earth is $\Delta t_{back} = 0$. Therefore, as soon as the spaceship reaches the planet it arrives back on Earth. To be clear, what returns to Earth immediately is not only an image of the spaceship as perceived on Earth, but the actual physical spaceship itself.

\begin{figure}[h]
\includegraphics[angle=90, width=18cm]{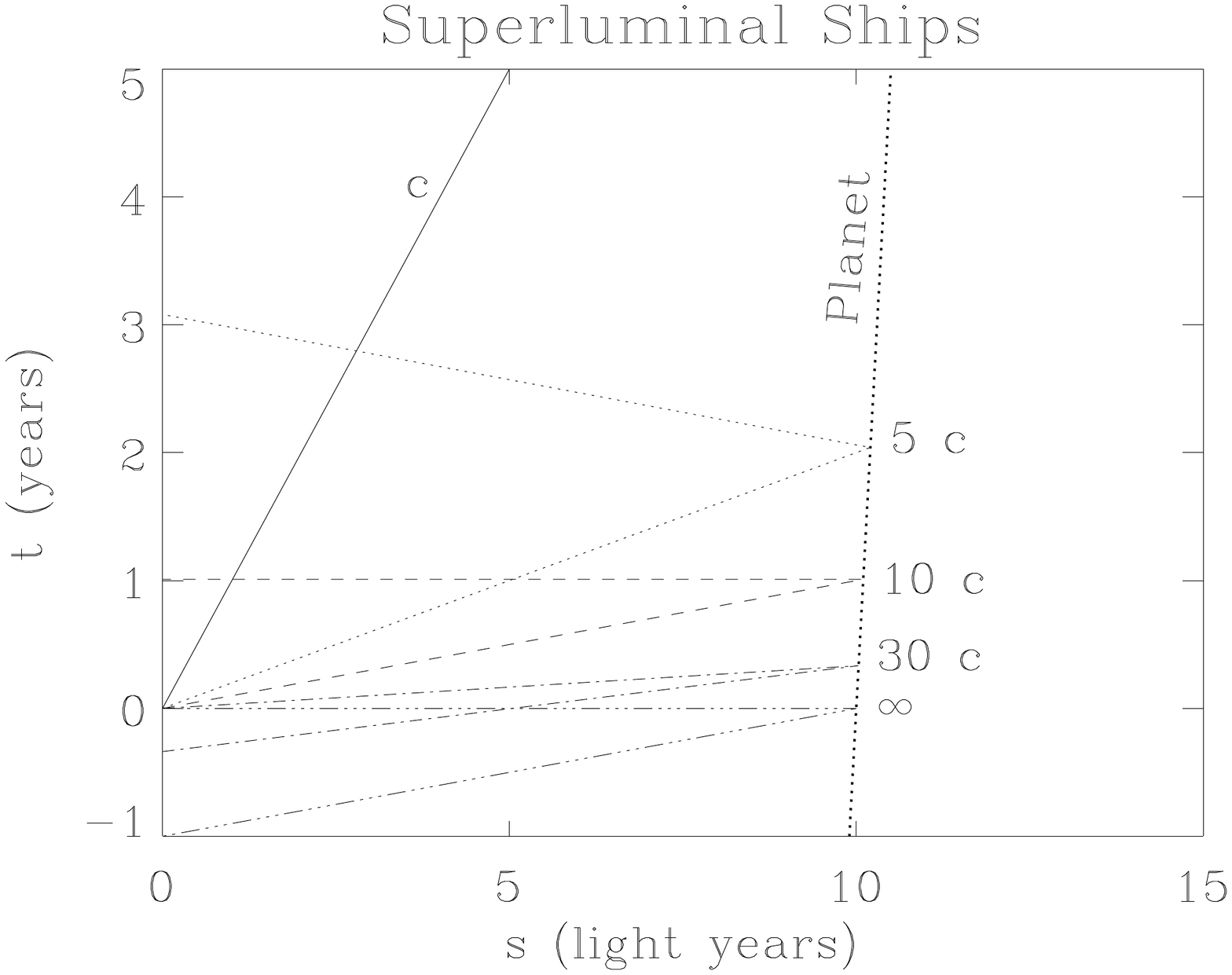}
\caption{A spacetime diagram is shown for the journeys taken by the superluminal ships described in the text. Time is plotted against distance traveled, both measured in the frame of the Earth. Earth remains at $(0,0)$. The world line of the planet continually receding from the Earth at $v_p = 0.1 c$, in the Earth frame, is depicted by the dark dotted line on the right. A ship traveling out at $v = c$ to the planet is shown by the solid line. Spaceships traveling out from the Earth at speeds of ($v = 5 c$, $10 c$, $30 c$, and $\infty$) are depicted by the (dotted, dashed, dot-dashed, triple dot-dashed) world lines respectively. World lines of similar type that connect back to Earth at times other than $t = 0$ are formally described as returning to Earth at the designated speed in the frame of the planet. However, as detailed in the paper, world lines that ``return" to Earth at $t < 0$ are actually time-reversed and described by the formalism as heading out from the Earth. Therefore, these world lines better describe the trip of a second ship that left Earth before the first ship, and arrived at the planet simultaneously with the first ship. All of these world lines depict actual spaceship positions and {\it not} the images of the ships as seen back on Earth.}
\label{superfast}
\end{figure}

In a specific example, the spaceship speed is now taken to be $v = c^2/u = 10 c$. This spaceship catches up with the planet at about $x_{turn} = 10.10$ light years at time $t_{out} = 1.010$ years after launch. With zero return time, Earth receives the spaceship back after $\Delta t_{tot} = 1.010$ years away. The dashed line in Figure \ref{superfast} describes the world line for this trip's journey.

In general, the Earth-bound observer sees the same series of events as perceived when $c < v < c^2/u$, they just happen a bit more compact in time. First, the spaceship is seen leaving. Next, a pair of spaceships appears on Earth on the Landing Pad. One ship from this image pair immediately leaves for the planet, while the other spaceship -- and its image -- remain on Earth. Both outbound spaceship images appear to reach the planet at the same time, and both then disappear from view.

In the specific example when $v = 10 c$, the Earth observer first sees the spaceship leave the Launch Pad at $t = 0$. At $t = 1.010$ years, the Earth observer suddenly sees two images appear on the Landing Pad, one of which immediately takes off -- time reversed -- toward the planet, while the other image stays put. At $t = 11.11$ years, the Earth observer sees both the outbound and return spacecraft images reach the planet, and both disappear.

\subsection{Spaceship speeds $c^2/u < v < c^2/u + c \sqrt{c^2/u^2 - 1}$ }

For spaceship velocities $c^2/u < v < c^2/u + c \sqrt{c^2/u^2 - 1}$, the return velocity $w$, in general, becomes formally positive. Since the spacecraft never moves toward the Earth in this scenario, how can it return to Earth? Although such a conundrum may seem like an end to a physically reasonable scenario, a physically consistent sequence of events does exist that is compatible with the formalism. This sequence is as follows. A spaceship leaves the Launch Pad on Earth for the planet. At a later time, as measured on Earth, a physical pair of spaceships materializes on the Landing Pad on Earth. One of these spaceships immediately goes off to the planet, while the other spaceship remains in place on the Landing Pad. Eventually both the spaceship that initially left the Launch Pad and the spaceship that later left the Landing Pad arrive at the planet simultaneously and dematerialize into nothing. 

There is a fundamental difference between this sequence of events and the events when $c < v \le c^2/u$. When the spacecraft has speeds in this range, a real pair creation event occurs on the Landing Pad, and a real pair annihilation event occurs at the planet. These are not images, but are consistent with the location(s) of the actual spacecraft(s) to any observer in the inertial frame of Earth, as computed by Eqs. (\ref{xout}) and (\ref{xback}). Interpreting these equations as describing physical spaceship pair events is a natural extension of the {\it image} pair events that occurs at lower speeds. Although observers in Earth's inertial frame that are located off the Earth may perceive events and sequences of events differently, they all must use the actual locations of the spacecraft as computed in the inertial frame of the Earth as the basis for what they see. 

In a specific example, the spaceship speed is now taken to be $v = 15 c$. The spaceship takes off from the Launch Pad at $t = 0$. Eq. (\ref{tout}) gives the outbound time as $\Delta t_{out} = 0.6711$ years and Eq. (\ref{tback}) gives the return time as $\Delta t_{back} = -0.3378$ years, so that the total time the spaceship is away from Earth is $\Delta t_{tot} = 0.3333$ years. Therefore the next thing that happens is that a pair of spaceships materialize on the Landing Pad at $t = 0.3333$ years. One spacecraft stays on the Landing Pad. The speed of the spaceship that leaves the Landing Pad is from Eq. (\ref{weq}) $w = 29.80 c$. Therefore, even though this ship left later, it is just the right amount faster to arrive at the planet at the same time as the spacecraft that left the Launch Pad. Both spacecraft catch up to the planet when it is, from Eq. (\ref{xturn}), $x_{turn} = 10.07$ light years distant. This occurs at is $t = 0.6711$ years. At this time, both outbound spacecraft merge and dematerialize.

A potential point of confusion is that the equation for the location of the spacecraft that left the Landing Pad, Eq. (\ref{xback}), formally returns a negative valued location for the spaceship when $t < 0.3333$ years. It is claimed here that such locations are outside of the described scenario and so do not occur. The two spaceships that materialize at $t = 0.3333$ years on the Landing Pad do not have previous positions described in this scenario. The situation is similar to the equation for the spaceship that left the Launch Pad, Eq. (\ref{xout}). This equation also does not indicate that the outbound spaceship that leaves the Launch Pad occupied negative valued locations before $t = 0$, as such positions are outside the described scenario and do not occur. 

Surprisingly, perhaps, what Earth observers see, in general, is {\it not} conceptually different from events perceived when the spacecraft has $c < v \le c^2/u$ as described in the previous two sections. Still the first event witnessed is the launching of the spacecraft from the Launch Pad. Next, Earth observers see a pair of spacecraft appear on the Landing Pad, one of which stays put and the other goes off to the planet. Last, the observers see both spacecraft arrive at the planet at the same time and disappear. 

In the specific example, a spaceship is seen from Earth to leave the Launch Pad at $t=0$. Suddenly, at $t = 0.3378$ years, a pair of spaceships appear on the Landing Pad, one of which stays there, and the other leaves for the planet. The spaceship that leaves the Landing Pad appears time reversed and faster than the spaceship that left Launch Pad. The images of the spacecraft arriving at the planet, described by Eq. (\ref{iout}) and Eq. (\ref{iback}), arrive back on Earth at $t = 10.74$ years, where the images appear to merge and disappear.

\subsection{Threshold Speed: $v = c^2 / u + c \sqrt{ c^2/u^2 - 1}$ }

The spacecraft speed $v = c^2 / u + c \sqrt{ c^2/u^2 - 1}$ is a threshold value because here $\Delta t_{back} = - \Delta t_{out}$, meaning that $\Delta t_{tot} = 0$ and that the total time it takes the spacecraft to go out to the planet and return back to Earth is zero. This speed is one of two formal solutions to setting $\Delta t_{tot}$, as given in Eq. ({\ref{ttot}), equal to zero. The other formal solution, $v = c^2 / u - c \sqrt{ c^2/u^2 - 1}$ always results in a $v < u$ and so is discarded because it describes a scenario where the spaceship never reaches the planet.

The general scenario has a spacecraft traveling in this speed range leave from the Launch Pad toward the planet. At the same time, a pair of spacecraft together materialize on the Landing Pad, one of which also immediately leaves for the planet, while the other spacecraft remains in place. The two simultaneously launched spacecraft both approach the planet, arrive simultaneously, and then de-materialize together at the planet. 

In a specific example, the spacecraft is assigned the speed $v = 19.95 c$. The return speed is computed from Eq. (\ref{weq}) to be, also, $w = 19.95 c$. Therefore, at $t = 0$, one spacecraft leaves the Launch Pad, while another spacecraft leaves the Landing Pad. A third spacecraft remains in place on the Landing Pad. Both outgoing spacecraft reach the planet at $t = 0.5038$ years when the planet is $x_{turn} = 10.05$ light years from Earth. At this time, both spacecraft merge and dematerialize.

What is perceived as happening for spacecraft in this general speed range is qualitatively the same as what happens. Earth observers see a spacecraft image launch for the planet from the Launch Pad and, simultaneously, a second spacecraft image leave for the planet from the Landing Pad. The image of the other member of the spacecraft pair that appears on the Landing Pad stays put. Images of both outgoing spacecraft are seen to approach the planet, arrive at the planet at the same time, and disappear together when they reach the planet.

In the specific example of $v = 19.95 c$, Earth observers see an image of the spacecraft leave from the Launch Pad -- and another image leave from the Landing Pad -- at $t = 0$. Images of both spaceships are seen to arrive at the planet at $t = 10.55$ years, whereafter the images merge and disappear.

\subsection{Spaceship speeds $v > c^2 / u + c \sqrt{ c^2/u^2 - 1}$ }

Cases where faster-than-light motion leads to closed-loop backward time travel can finally be explained as a logical extension of previously discussed results. In cases of increasing $v$ where $v > c^2 / u + c \sqrt{ c^2/u^2 - 1}$, in general, the ``return back" speed $w$ is not only positive -- and so indicating motion away from the Earth -- but decreasing -- and so indicating slower motion toward the planet. The result is that the spacecraft ``returns back" to the Landing Pad {\it before} it leaves from the Launch Pad. Surprisingly, this scenario does not give the straightforward out-and-back sequence of events commonly assumed for superluminal time-travel. On the contrary, this scenario relies on pair-creation and pair-annihilation events. 

The first scenario event for spacecraft speeds in this general speed range is that two spacecraft appear on the Landing Pad, one of which immediately sets off for the planet. The next event, as described in the Earth frame, is that the initial spacecraft takes off from the Launch Pad and heads out toward the planet. Both the spacecraft that left from the Landing Pad and the spaceship that left from the Launch Pad reach the planet at the same time and de-materialize. 

In the specific example of $v = 30 c$, Eq. (\ref{weq}) now yields a ``return back" speed of $w = 14.95 c$, a positive value that describes movement away from the Earth. Note also that $w < v$, so that $w$ describes less rapid outward motion. Further, Eq. (\ref{tout}) shows $\Delta t_{out} = 0.3344$ years, while Eq. (\ref{tback}) shows $\Delta t_{back} = -0.6711$ years, so that their sum has $\Delta t_{tot} = -0.3367$ years, meaning that the spaceship ``returns" 0.3367 years before it left. The dot-dashed line in Figure \ref{superfast} describes the world line for this trip's journey.

Eq. (\ref{xout}) describes the spaceship that left from the Launch Pad at $t = 0$. Eq. (\ref{xturn}) shows that this $v = 30 c$ spacecraft catches up with the planet at $x_{turn} = 10.03$ light years from Earth. The time this spaceship catches up to the planet is at $t_{out} = 0.3344$ years. 

Eq. (\ref{xback}) describes the spaceship that left from the Landing Pad at $t = -0.3367$ years. This spacecraft also catches up to the planet at $t = 0.3344$ years. Times greater than $t = 0.3344$ years yield distance values in Eq. (\ref{xback}) larger than that of the planet, but these are considered unphysical because the scenario gives the boundary condition that the spacecraft turns around at the planet. Similarly, times less than $t = -0.3367$ years yield negative distance values in Eq. (\ref{xback}), but these are also considered unphysical because the scenario states a boundary condition that the spacecraft travels only between Earth and the planet. One might argue that all times before $t = 0$ are similarly unphysical because the scenario dictated that the spaceship launched at $t = 0$, but no such temporal boundary condition was placed on the time of return. 

It is educational to query the locations of the $v = 30c$ outbound and ``return back" spaceships during their journeys to see how they progress. At time $t = -0.3367$ years, Eq. (\ref{xback}) indicates that the return spaceship has a location of $x_{back} (-0.3367) = 0$ light years, meaning the ``return back" spaceship is still on Earth. At time $t = -0.1$ years, this equation indicates that the return spaceship has a location $x_{back} = 3.538$ light years from Earth in the direction of the planet. At time $t = 0$ years, Eq. (\ref{xout}) indicates that $x_{out} (0 \ \rm{years}) = 0$ light years, meaning that the ``outbound" spaceship is still on Earth, while Eq. (\ref{xback}) indicates that $x_{back} = 5.033$ light years from Earth. At time $t = 0.1$ years the equations hold that $x_{out} (0.1 \ \rm{years}) = 3.000$ light years, while $x_{back} (0.1 \ \rm{years}) = 6.528$ light years. Next, at $t = 0.2$ years, the equations hold that $x_{out} (0.2 \ \rm{years}) = 6.000$ light years, while $x_{back} (0.2 \ \rm{years}) = 8.023$ light years. Finally, at $t = 0.3344$ years, both $x_{out} (0.3344 \ \rm{years}$) and $x_{back} \ \rm{years}$ yield 10.03 light years. 

What is seen on Earth in general for speeds in this range is again qualitatively similar to what physically happens. First, two spacecraft images are seen to appear on the Landing Pad, one of which is seen to launch immediately for the planet, while the other appears to stay put. Next, an image of the spacecraft on the Launch Pad launches for the planet. Both the image of the spacecraft that left from the Landing Pad and the image of the spacecraft that left from the Launch Pad appear to reach the planet at the same time. These two images merge and disappear.

In the specific example where $v = 30 c$, Earth sees the spaceships that materialized on the Landing Pad at $t = -0.3367$ years with no time delay because this Landing Pad is on Earth. The spaceship that stays on the Landing Pad is always seen on Earth to be on the Landing Pad with no time delay, from $t = -0.3367$ years onward. The spaceship that left the Landing Pad for the planet at $t = -0.3367$ is seen with increasing time delay due to the travel time of the light between the spaceship and Earth. At $t = -0.1$ years, the ``return back" spaceship is at $x_{back} = 3.538$ light years from Earth but because of light travel time, is seen when it was only at $x_{back} = 0.2218$ light years away. At $t = 0$ years, a spaceship is seen to take off from the Launch Pad, while the spaceship that left the Landing Pad while actually at $x_{back} (0) = 5.033$ light years distant, appears as it did when at $x_{back} = 0.3156$ light years distant. At $t = 0.2$ years $x_{out} (0.2 \ \rm{years})$ = 6.000 light years, but due to light travel time, this ``outbound" spaceship appears as it did when at $x_{out} = 0.1935$ light years. Similarly, at $t = 0.2$ years, the ``return back" spaceship is at $x_{back} (0.2 \ \rm{years}) = 8.023$ light years distant, but due to light travel time appears as it did when at $x_{back} = 0.5030$ light years. Images of both spaceships arriving at the planet are received back on Earth at $t = 10.367$ years. At this time, both images merge and disappear.

Even before the outbound spaceship leaves from the Launch Pad, astronauts on the spaceship that appeared and remained on the Landing Pad may come out, recount their journey, and even watch the subsequent launch of their spacecraft on the nearby Launch Pad. A physical conundrum occurs, for example, if these astronauts go over to the Launch Pad and successfully interfere with the initial spacecraft launch. This would create a causal paradox that may reveal any time travel to the past to be unphysical. \citep{Hossenfelder2012} Alternatively, however, the presented scenario may proceed but the disruption event may be disallowed by the Novikov Chronology Protection Conjecture \citep{Novikov1992} -- or a similarly acting physical principle. Then, try as they might, the astronauts could find that they just cannot disrupt the launch. \citep{Lloyd2011} In a different alternative, such disruption actions may be allowed were the universe to break into a sufficiently defined multiverse, with the disruption just occurring in a different branch of the multiverse \citep{Everett1957} than the one where the spacecraft initially launched. Then, life for the disrupting astronauts would continue on normally even after they disrupted the launch, even though they could remember this launch. It is not the purpose of this work, however, to review or debate causal paradoxes created by backward time travel, but rather to show how some superluminal speeds do not lead to closed-loop backward time-travel, while other speeds do, but by incorporating non-intuitive pair creation and annihilation events.

\subsection{When $v$ Diverges}

Perhaps counter-intuitively, an infinite amount of backward time travel does not result when $v$ diverges. In the general case, as $v$ approaches infinity, the time it takes for the spacecraft to reach the planet, $\Delta t_{out}$, approaches zero. However, Eq. (\ref{tback}) shows that the time it takes for the spacecraft to return to Earth, $\Delta t_{back}$, does {\it not} approach negative infinity but rather $\Delta t_{back} \sim - x_{po} u / c^2$. The reason for this is that, in Eq. (\ref{weq}), $w$ only approaches $c^2/u$ as $v$ diverges, which is always superluminal and never zero. Therefore, the faster the planet is moving away from the Earth, and the further the planet is initially from the Earth, the further back in time the returning spaceship may appear on the Landing Pad. 

In the specific example, diverging $v$ leads to a maximum backward time travel according to Eq. (\ref{tback}) of $\Delta t_{back} = 1.000$ year. The return speed $w$ approaches 10 c. Therefore, in this scenario, the earliest a spaceship pair could appear on the Landing Pad would be $t = -1.000$ year, one year before the outbound spaceship leaves the Launch Pad. After this, events would unfold qualitatively as described in the last section. The triple dot-dashed line in Figure \ref{superfast} describes the world line for this trip's journey.

\section{PLANET-FREE SCENARIOS}

One might consider that the pair creation and annihilation arguments above only arise because of the ``trick" of involving a planet that has a non-zero and positive speed $u$. In this view the planet's speed, along with the relativistic speed addition formula, act as a spurious door to mathematical possibilities that are physically absurd. As evidence, one might take an example where a spaceship leaves with a speed $v$ relative to the Earth and then returns at a speed $u$, again relative to the Earth. The arbitrary turnaround location can be labeled $x_{turn}$. Then, in the Earth frame, the time it takes for the ship to reach the turnaround location would be $\Delta t_{out} = x_{turn} / v$, and the time it takes for the ship to return from the turnaround location would be the same: $\Delta t_{back} = x_{turn} / u$. When $u > c$, a pair of virtual images of the ship will again be seen, for a while, on Earth. However, there are no velocities $v$ and $u$, subluminal or superluminal, where either $\Delta t_{out}$ or $\Delta t_{back}$ is negative, and so no $v$ and $u$ values exist that create $\Delta t_{tot} < 0$. Therefore, in this scenario the spaceship will never arrive back on Earth before it left. Does this counter-example disprove the presented analysis?

No. The scenario of the previous paragraph does not create a situation where an object returns to the same location at an earlier time -- a closed-loop backward time travel event. Therefore this scenario does not address the main query posed by this work -- how faster-than-light travel enables backward time travel.

Scenarios do exist, however, where superluminal travel creates closed-loop backward time travel events, but where no intermediary planet is involved. Such scenarios, which some might consider simpler, have the spaceship just go out at one speed, turn around at an arbitrary location, and return at another speed. So long as the return speed is generated and hence specified relative to the outbound speed, then the relativistic velocity addition formula Eq. (\ref{weq}) may be used, and the same types of results arise. Note that is only presumed that Eq. (\ref{weq}) is valid when one or both speeds $v$ and $u$ are superluminal -- this presumption has never been verified.  

A simple planet-free scenario is as follows. A spaceship leaves Earth at speed $v$. At an arbitrary turnaround location, the ship changes its velocity by $2 v$, toward the Earth, relative to its outward motion. For non-relativistic speeds, this turnaround would result in the ship heading toward Earth at speed $v$. For relativistic and superluminal speeds, however, the relativistic velocity addition formula must be used, resulting in more complicated behavior. Specifically, in this scenario, it is straightforward to show following the above logic that $\Delta t_{tot} = 2 \Delta t_{out} (1 - v^2/c^2)$, so that closed-loop backward time travel events occur for all spaceship speeds of $v > c$. As before, tracking spacecraft and image locations show that pair events also may occur.

This brings up the question: why does it matter against what the spaceship's relative return velocity is measured -- shouldn't the physics be the same? Coordinate invariance -- called general covariance, and inertial frame invariance -- called Lorentz invariance -- should make the physics the same no matter which coordinates are used for tracking and no matter which inertial frames are used for comparison. Specifically, in this case, closed-loop backward time travel should not depend on whether the spaceship's return velocity is specified relative to the Earth, or a planet, or the spaceship's previous velocity, or anything else. The reality of what happens should be same regardless. 

The key symmetry-breaking point is that the standard special relativistic velocity transformation, Eq. (\ref{weq}), is not confined to be Lorentz invariant when both subluminal and superluminal speeds are input. Mathematically, the reason is that the denominator of the velocity addition formula goes through a singularity at $u v = c^2$, a singularity that cannot be reversed by a simple coordinate or inertial frame transformation. Physically, turning around relative to a different object may change the scenario -- a different physical process may be described.

\section{DISCUSSION AND CONCLUSIONS} 

An analysis has been given showing how faster-than-light travel can result in closed-loop backward time travel. The analysis focused on an extremely simple scenario -- an object going out and coming back -- effectively extending the twin paradox scenario to superluminal speeds. Further, only a single relativistic formula was used -- that for velocity addition. A surprising result is that, in this scenario, backward time travel appears only when the turnaround location is moving away from the launch location, and, further, is bound to the creation and annihilation of object pairs. The underlying mathematical reason is that the negative time duration for the {\it return} trip needed to create closed-loop backward time travel is tied to spaceship motion {\it away} from the launch site, not toward it, as shown in Eq. (\ref{tback0}). This behavior neatly describes an (Earth-observed) spaceship moving out toward the planet on the ``return back" leg of the trip in addition to the (Earth observed) spaceship moving out toward the planet on the initial ``outward" leg of the trip. One knows that the spaceship does return, and so the second member of the pair-created spaceships remains on the Landing Pad. Note that when superluminal spaceship speeds are invoked, the spaceship {\it always} travels at superluminal speeds relative to the Earth, and never accelerates through $c$. 

It is tempting to explain away these results as meaningless because the relativistic velocity addition formula, Eq. (\ref{weq}) was applied to a regime where it might not hold: where one speed is superluminal. However, the validity of this formula in the superluminal regime should be testable in a conventional physics lab where illumination fronts or sweeping spots move superluminally, in contrast to a detector that moves subluminally. \cite{Nemiroff2016} Further, to our knowledge, no other relativistic velocity addition formula has even been published.

Although not defined in the above equations, it is consistent to conjecture that the superluminal spaceship has negative energy. \citep{Chase1993} This may be pleasing from an energy conservation standpoint because both pair creation and annihilation events always involve a single positive energy and a single negative energy spaceship -- never two positive energy or two negative energy spaceships. Therefore, neither the creation nor annihilation of a spaceship pair, by themselves, demand that new energy be created or destroyed. 

It is not clear how ``real" the negative energy spaceships are to observers in inertial frames other than the Earth, including frames moving superluminally. The negative energy ships are surely real in the Earth frame in the sense that they give those observers positions from which spaceship images can emerge. However, these negative energy ships may not exist in some other reference frames, which appears to raise some unexplored paradoxes. Also unresolved presently is whether observer in a superluminal positive-energy ship that left the Launch Pad would be able to see a negative-energy ship that left from the Landing Pad. Since it is not in the scope of the above work to analyze what happens in inertial frames other than the Earth, then, unfortunately, this and other intriguing questions will remain, for now, unanswered.

Finally, this analysis may give some unexpected insight to physical scenarios that seem to depend on superluminal behavior. For example, implied non-locality in quantum entanglement typically posits some sort of limited superluminal connection between entangled particles, although one that does not allow for explicit superluminal communication. To the best of our knowledge, never has such supposed superluminal connection been tied through the special relativity addition formula to pair events. Perhaps one reason for this is that so few seem to know about it. Yet, as implied here -- it may well be expected for observers in some reference frames. 

\acknowledgments

The authors thank Qi Zhong, Teresa Wilson, and Chad Brisbois, for helpful conversations.

\clearpage

\begin{table}[h]
\begin{center}
\begin{tabular}{| c | c | c | c | c | c |}
\hline
\vspace{-0.13in}
$v$ & $w$ & $x_{turn}$ & $\Delta t_{out}$ & $\Delta t_{back}$ & $\Delta t_{tot}$  \\
($c$) & ($c$) & (light years) & (years) & (years) & (years) \\
\hline 
0.1 & -- & $\infty$ & $\infty$ & -- & -- \\
\hline
0.5 & -0.4210 & 12.50 & 25.00 & 29.69 & 54.68 \\
\hline
1.0 & -1.000 & 11.11 & 11.11 & 11.11 & 22.22 \\
\hline
5.0 & -9.800 & 10.20 & 2.041 & 1.041 & 3.082 \\
\hline
10.0 & $\pm$ $\infty$ & 10.10 & 1.010 & 0.000 & 1.010 \\
\hline
15.0 & 29.80 & 10.07 & 0.6711 & -0.3378 & 0.3333 \\
\hline
19.95 & 19.95 & 10.05 & 0.5038 & -0.5038 & 0.0000 \\
\hline
30.0 & 14.95 & 10.03 & 0.3344 & -0.6711 & -0.3367 \\
\hline
$\infty$ & 10.00 & 10.00 & 0.000 & -1.000 & -1.000 \\
\hline
\end{tabular}
\caption{\label{table1}Time of travel parameters for the scenario given where a spaceship travels out to a planet at constant speed $v$, relative to the Earth, and returns back at constant speed $v$, relative to the planet. In this example the planet has initial distance from Earth of $x_{po} = 10$ light years and constant speed $u = 0.1 c$ away from the Earth.}
\end{center}
\end{table}

\end{document}